\documentclass[reprint,preprintnumbers,amsmath,amssymb,superscriptaddress,twocolumn]{revtex4}

\usepackage{graphicx}
\usepackage{dcolumn}
\usepackage{bm}

\begin{document}

\newcommand{\CMO}{CaMoO$_4$}
\newcommand{\enCMO}{$^{40}$Ca$^{100}$MoO$_4$}
\newcommand{\dbd}{$0\nu\beta\beta$}
\newcommand{\twodbd}{$2\nu\beta\beta$}
\newcommand{\mbb}{$m_{\beta\beta}$}

\preprint{}

\title{A {\CMO} Crystal Low Temperature Detector for the AMoRE Neutrinoless Double Beta Decay Search}

\author{G.B.~Kim}
\affiliation{Institute for Basic Science, Daejeon 305-811, Republic of Korea}
\affiliation{Department of Physics and Astronomy, Seoul National University, Seoul 151-747, Republic of Korea}
\affiliation{Korea Research Institute of Standards and Science, Daejeon 305-340, Republic of Korea}
\author{S.~Choi}
\affiliation{Department of Physics and Astronomy, Seoul National University, Seoul 151-747, Republic of Korea}
\author{F.A.~Danevich}
\affiliation{Institute for Nuclear Research, Kyiv 03680, Ukraine}
\author{A. Fleischmann}
\affiliation{Kirchhoff-Institut f\"ur Physik, Universit\"at Heidelberg, D-69120, Heidelberg, Germany}
\author{C.S.~Kang}
\affiliation{Institute for Basic Science, Daejeon 305-811, Republic of Korea}
\affiliation{Korea Research Institute of Standards and Science, Daejeon 305-340, Republic of Korea}
\author{H.J.~Kim}
\affiliation{Physics Department, Kyungpook National University, Daegu 702-701, Republic of Korea}
\author{S.R.~Kim}
\affiliation{Institute for Basic Science, Daejeon 305-811, Republic of Korea}
\affiliation{Korea Research Institute of Standards and Science, Daejeon 305-340, Republic of Korea}
\author{Y.D.~Kim}
\affiliation{Institute for Basic Science, Daejeon 305-811, Republic of Korea}
\affiliation{Department of Physics, Sejong University, Seoul 143-747, Republic of Korea}
\author{Y.H.~Kim}
\email{yhkim@kriss.re.kr}
\affiliation{Institute for Basic Science, Daejeon 305-811, Republic of Korea}
\affiliation{Korea Research Institute of Standards and Science, Daejeon 305-340, Republic of Korea}
\affiliation{Korea University of Science and Technology, Daejeon 305-350, Republic of Korea}
\author{V.A.~Kornoukhov}
\affiliation{Institute for Theoretical and Experimental Physics, Moscow 117218, Russia}
\author{H.J.~Lee}
\affiliation{Institute for Basic Science, Daejeon 305-811, Republic of Korea}
\affiliation{Korea Research Institute of Standards and Science, Daejeon 305-340, Republic of Korea}
\author{J.H.~Lee}
\affiliation{Korea Research Institute of Standards and Science, Daejeon 305-340, Republic of Korea}
\author{M.K.~Lee}
\affiliation{Korea Research Institute of Standards and Science, Daejeon 305-340, Republic of Korea}
\author{S.J.~Lee}
\altaffiliation{Now at NASA Goddard Space Flight Center, Greenbelt, MD 20771, USA}
\affiliation{Korea Research Institute of Standards and Science, Daejeon 305-340, Republic of Korea}
\author{J.H.~So}
\affiliation{Institute for Basic Science, Daejeon 305-811, Republic of Korea}
\affiliation{Korea Research Institute of Standards and Science, Daejeon 305-340, Republic of Korea}
\author{W.S.~Yoon}
\affiliation{Institute for Basic Science, Daejeon 305-811, Republic of Korea}
\affiliation{Korea Research Institute of Standards and Science, Daejeon 305-340, Republic of Korea}
\affiliation{Korea University of Science and Technology, Daejeon 305-350, Republic of Korea}

\begin{abstract}
We report the development of a \CMO{} crystal low temperature detector for the AMoRE neutrinoless double beta decay (\dbd) search experiment. 
The prototype detector cell was composed of a 216 g \CMO{} crystal and a metallic magnetic calorimeter. An over-ground measurement demonstrated FWHM resolution of 6-11~keV for full absorption gamma peaks. Pulse shape discrimination was clearly demonstrated in the phonon signals, and 7.6~$\sigma$ of discrimination power was found for the $\alpha$ and $\beta / \gamma$ separation. 
The phonon signals showed rise-times of about 1~ms. It is expected that the relatively fast rise-time will increase the rejection efficiency of two-neutrino double beta decay pile-up events which can be one of the major background sources in {\dbd} searches.

\begin{description}

\item[Keyword]{Double beta decay, low temperature detector, \CMO{} crystal scintillator, particle discrimination}
\end{description}
\end{abstract}

\pacs{Valid PACS appear here}
\maketitle

\section{Introduction}
Recent neutrino oscillation experiments have been unveiling the properties of neutrinos~\cite{rev_pp,Mohapatra_theory_neu}. 
Their experimental evidences strongly suggest that neutrinos are massive and encounter flavor mixing of mass eigenvalues. The mixing angles and the differences between the square masses have been estimated.
However, those observations do not provide a direct measurement of the absolute mass, and do not answer the question of whether neutrino is its own anti-particle (Majorana-type) or not (Dirac-type). 

Search for neutrinoless double beta  decay (\dbd) is a key experiment to reveal un-answered nature of neutrinos~\cite{elliott_dbd,Avignone_dbd,rodejohann_dbd,andrea_ahep,gomez2012search}. 
The double beta decay (\twodbd) that accompanies the simultaneous emission of two electrons and two anti-neutrinos is a rare process that is an allowed transition in the standard model. 
Another type of double beta decay that does not emit any neutrinos, \dbd, can occur if neutrino is massive Majorana particles (i.e., it is its own anti-particle). In the \dbd{} process, the full available energy of the decay is carried by the two electrons and the recoiled daughter which has a very small amount of energy compared with that of the electrons. Therefore, while the electron sum energy spectrum in the \twodbd{} is continuous up to the available energy release ($Q_{\beta\beta}$), in the \dbd{}, the spectrum should have a sharp peak at $Q_{\beta\beta}$~\cite{Mohapatra_theory_neu}.

The observation of {\dbd} would clearly demonstrate that neutrino is not Dirac-type but rather is Majorana-type particle. In that case, physical processes that do not conserve lepton number would be allowed. Moreover, the absolute mass scale, so-called ``effective Majorana neutrino mass'', $\langle m_\nu \rangle$ can be estimated using the \dbd{} half life $(T^{0\nu}_{1/2})$,
\begin{equation}
\centering{
\left( T^{0\nu}_{1/2} \right)^{-1}  = G^{0\nu} |M^{0\nu}|^2  \frac {\langle m_\nu \rangle^2 }{m_e^2} 
}
\end{equation} 
where $m_e$ is the electron mass, $G^{0\nu}$ is the kinematic phase-space factor calculable with reasonable precision, and $M^{0\nu}$ is the model dependent nuclear matrix element. Here, the  Majorana mass is defined as 
\begin{equation}
\centering{
\langle m_\nu \rangle = \left| \sum{U^2_{ej}m_j} \right|
}
\end{equation}
where the $m_j$'s are the mass eigenstates of the neutrino, $U_{ej}$'s are the elements of the mixing matrix between the flavor states and mass eigenstates.  

Experimentally, the measurement limit of half life is often used as the sensitivity to probe the rare event~\cite{gomez2012search}.  In a measurement with non-negligible backgrounds, the sensitivity becomes
\begin{equation}
\centering{
T^{0\nu}_{1/2} \propto \delta \varepsilon \sqrt{ \frac{M t}{b \Delta E}} 
}
\end{equation}
where $\delta$ is the concentration of $\beta\beta$ isotope in the detector, $\varepsilon$ is the detection efficiency, $M$ is the detector mass, $t$ is the measurement time, $b$ is the background rate per unit mass and energy, and $\Delta E$ is the energy resolution of the detector, in other words, the region of interest (ROI) of the energy window at the $Q_{\beta\beta}$ value. However, in a case of a zero-background experiment that observes no event in ROI during the measurement time, the sensitivity becomes proportional to the detector mass and the measurement time,
\begin{equation}
\centering{
 T^{0\nu}_{1/2} \propto \delta \varepsilon M t.
}
\end{equation}

\begin{figure}
        \centering
        \includegraphics[width=1\linewidth,keepaspectratio]{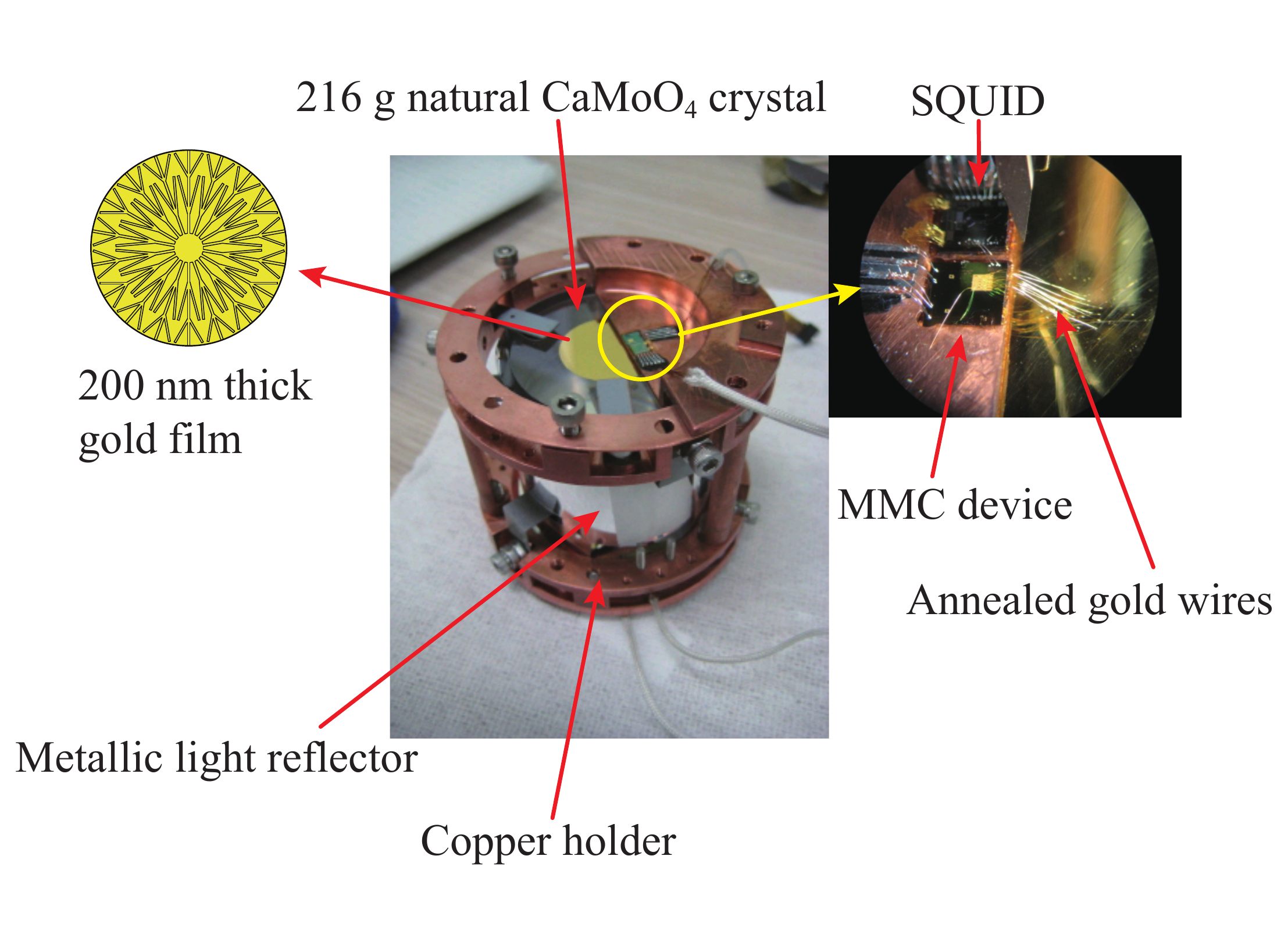}
    \caption{(Color online) Low temperature detector setup with a 216 g {\CMO} crystal and an MMC sensor.}
        \label{fig:detector_design}
\end{figure}

To increase the detection sensitivity, it is essential to have a detector with high concentration of the isotope of interest, detection efficiency, energy resolution and efficient background rejection capability as well as to minimize backgrounds from internal and external sources in the region of interest. 
High energy resolution and detection efficiency experiment can be realized with crystal detectors containing the isotope of interest.
The detector performance of recently developed low temperature detectors (LTDs) that operate at sub-Kelvin temperatures can perfectly meet the requirements by utilizing state-of-the-art detector technologies with extreme energy sensitivity, such as neutron transmutation doped (NTD) Ge thermistors, superconducting transition edge sensors (TESs), or metallic magnetic calorimeters (MMCs) \cite{enss2005cryogenic}.   

The AMoRE (Advanced Mo-based Rare process Experiment) project is an experiment to search for {\dbd} of $^{100}$Mo~\cite{bhang_AMoRE}. AMoRE uses {\CMO} crystals as the absorber and MMCs as the sensor~\cite{sjlee_app, gbkim_ltd15}. {\CMO} is a scintillating crystal that has the highest light output at room and low temperatures among Mo-containing crystals (molybdates)~\cite{hjkim_IEEE_2010,pirro_2006}.

Choosing of $^{100}$Mo as the source of \dbd{} is advantageous. The nucleus has a high Q value of 3034.40(17) keV~\cite{Rahaman2008111} that is above the intensive 2615 keV gamma quanta from $^{208}$Tl decay ($^{232}$Th family). The natural abundance of $^{100}$Mo is 9.8\%~\cite{Mo_abundance}, which is comparatively high.
Furthermore, enriched $^{100}$Mo can be produced by centrifugation method in amount of tens of kilograms per year with a reasonable price. Also, the theoretically estimated half life of $^{100}$Mo is relatively shorter than that of other \dbd{} candidates~\cite{barea2012_prl_109_042501_2012,Vergados_2012}. However, \twodbd{} of $^{48}$Ca with $Q_{\beta\beta}=4272$ keV (despite rather low concentration of the isotope $\approx0.2\%$) can be an irremovable background source in the ROI of $^{100}$Mo. The AMoRE collaboration has successfully grown \enCMO{} crystals using $^{100}$Mo enriched and $^{48}$Ca depleted materials. Three of the doubly-enriched crystals with masses in the range of 0.2-0.4~kg were tested in a low background 4$\pi$ veto system to determine their internal backgrounds~\cite{jhso_CMO_background}.

The present experimental work aims to test the low temperature detection concept with a \CMO{} crystal and an MMC that is suitable for a high resolution experiment to search for \dbd{} of $^{100}$Mo. A 216 g natural \CMO{} crystal with an MMC phonon sensor  was employed in this experiment, which was performed in an over-ground measurement facility. The energy resolution and linearity of the detector setup, particle and randomly coinciding events discrimination by pulse shape analysis for background rejection in a \dbd{} experiment are discussed in this report.

\section{Experimental Details}

\begin{figure}
        \centering
        \includegraphics[width=1\linewidth,keepaspectratio]{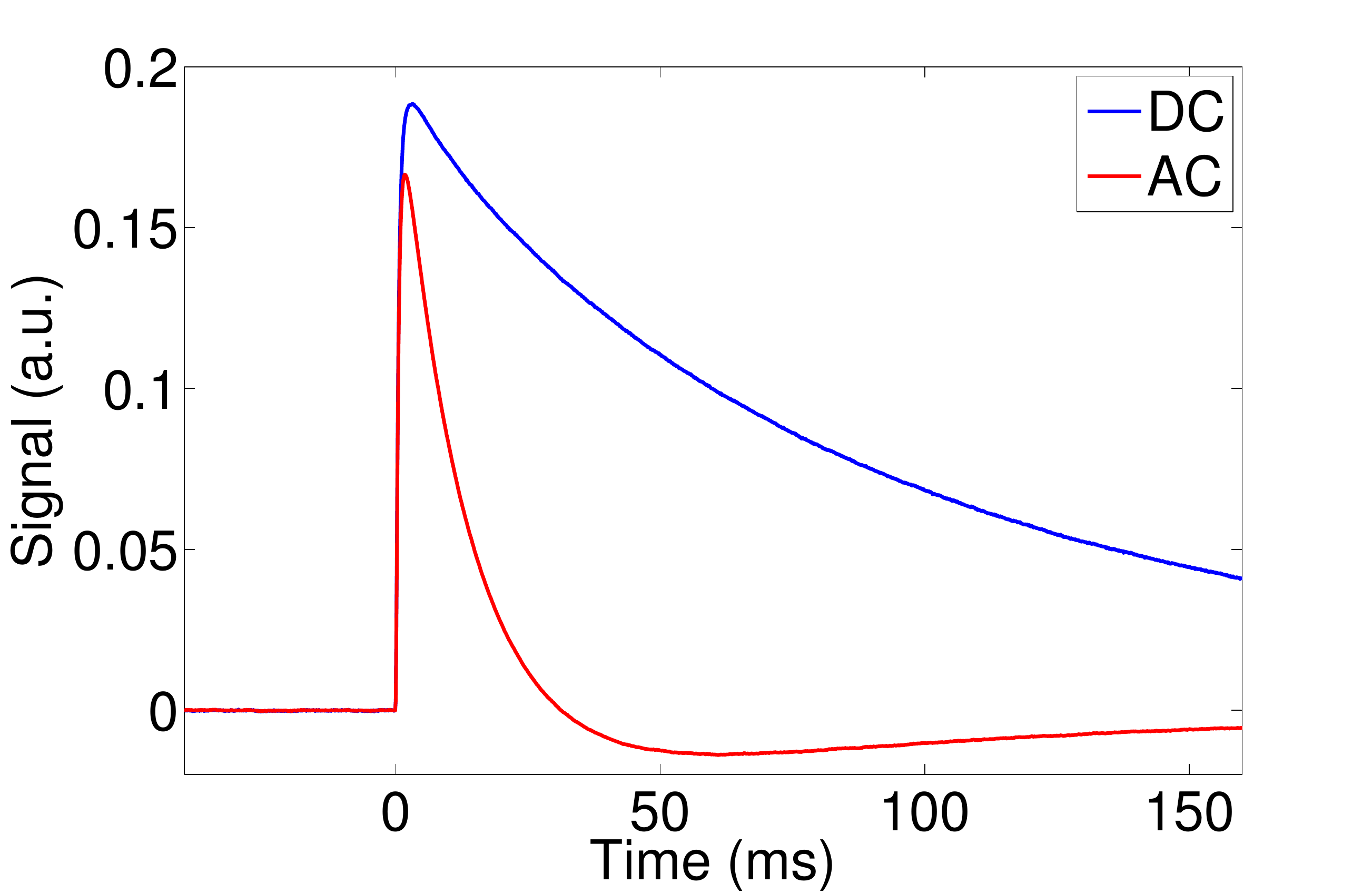}
    \caption{(Color online) A typical signal of 2.6 MeV gamma-ray full absorption events at 40 mK in DC (blue, bigger) and AC (red, smaller) coupling.}
        \label{fig:Tl_single}
\end{figure}
The detector setup was structured in a cylindrical shape with copper support details as shown in Fig.~\ref{fig:detector_design}. A \CMO{} crystal with 4 cm in diameter and 4 cm height was mounted inside the copper structure using metal springs. The mass of the crystal  was 216 g. It was grown with natural Ca and Mo elements at the Bogoroditsk plant in Russia. A patterned gold film was evaporated on one side of the crystal to serve as a phonon collector. An MMC device, the primary sensor for detecting the phonon signals absorbed in the gold film, was placed on a semi-circular copper plate over the crystal. The thermal connection between the gold film and the MMC was made using annealed gold wires. Details regarding the measurement principle and the detector structure of the MMC device were presented in previous reports \cite{wsyoon_ltd14, wsyoon_ltd15}.

When a particle hits a dielectric material, most of the energy deposited into the absorber is converted into the form of phonons. High energy phonons with frequencies that are close to the Debye frequency are generated initially. However, they quickly decay to lower frequency phonons via anharmonic processes. When their energy becomes 20-50 K, they can travel ballistically in the crystal \cite{wolfe}. The major down-conversion processes of these athermal phonons are isotope scattering, inelastic scatterings by impurities and lattice dislocations, and inelastic scatterings at crystal surfaces \cite{leman}. These excess phonons eventually change the equilibrium thermal phonon distribution, thereby causing temperature increase.  

In the detector setup with the \CMO{} crystal and the gold phonon collector film, the ballistic athermal phonons can hit the crystal and gold interface, transmit into the gold film, and transfer their energy to the electrons in the film \cite{yhkim2004}. The electron temperature of the gold film increases quickly via electron-electron scatterings. This temperature change is measured by the MMC sensor that is thermally connected with the gold wires. The size of the gold film and number of gold wires were chosen based on a thermal model study that considered the efficient athermal heat flow process~\cite{gbkim_ltd15}. Consequently the gold film had a diameter of 2 cm, a thickness of 200 nm, and an additional gold pattern of 200 nm thickness on top of the gold film. to increase the lateral thermal conductivity of the gold film. 

The detector assembly was installed in a dilution refrigerator in an over-ground laboratory at KRISS (Korea Research Institute of Standards and Science). The refrigerator was surrounded by a 10 cm thick lead shield (except the top surface) to reduce environmental gamma ray background. 
\begin{figure}
        \centering
        \includegraphics[width=1\linewidth,keepaspectratio]{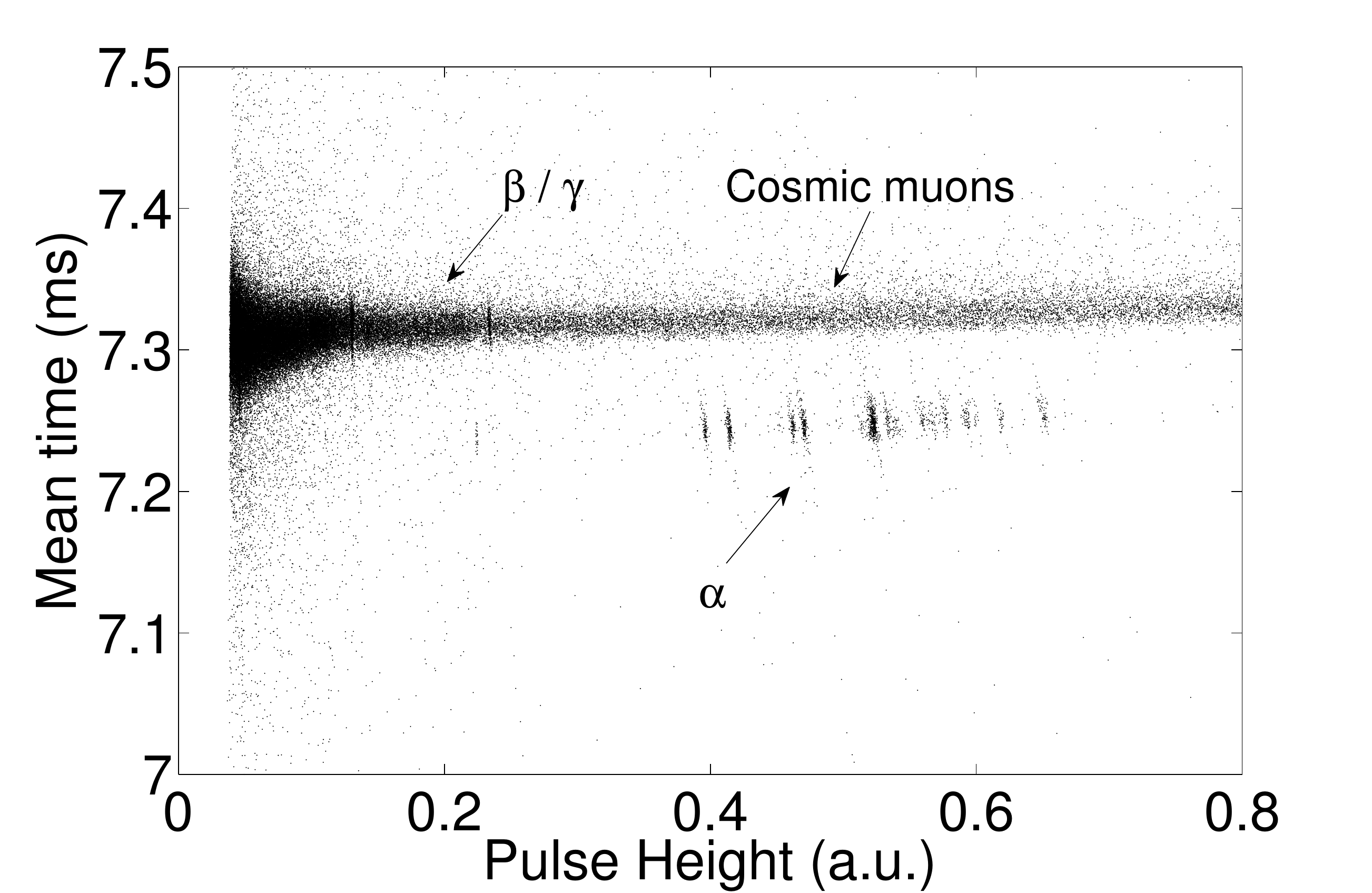}
    \caption{A scatter plot of the mean-time and pulse height obtained from the background measurement of 95 h in an over-ground laboratory. $\alpha$ and $\beta / \gamma$ (including cosmic muons) events are clearly separated in terms of their mean-time values.}
        \label{fig:PH_meantime}
\end{figure}
The detector with an MMC operates well in the temperature range of 10-50 mK. The signal size increases at lower temperatures since the MMC sensitivity enhanced and the heat capacities decreased. 
However, the signals have slower rise and decay times at lower temperatures as thermal conductances become poorer. Larger signal size improves the energy threshold and baseline energy resolution of the detector. 
The energy resolution of the detector measured for particle absorption events can be worse than the baseline resolution because of any uncorrelated mechanism that affects the signal size and shape. Examples of such mechanisms include temperature fluctuations due to instrumental instability or frequent event rates, position dependence of signal shapes, or scintillation processes that are associates with phonon generations in an inhomogeneous way.
Therefore, larger signal size does not guarantee a better energy resolution at certain temperatures. At the present experimental condition including the background rate from cosmic muons and external gamma-rays, 40 mK was selected as the main measurement temperature. At this temperature, about 1 ms rise-time was obtained for the 2.6 MeV gamma line without degrading the energy resolution. A typical signal of 2.6~MeV gamma-ray full absorption events is shown in Fig.~\ref{fig:Tl_single}. The rise-time of the DC coupled signal is 1.1 ms, which is somewhat slower than that of earlier measurements for which shorter gold wires were used \cite{gbkim_ltd15}. 
\section{Pulse Shape Analysis}

\begin{figure}
	\centering
                \includegraphics[width=1\linewidth,keepaspectratio]{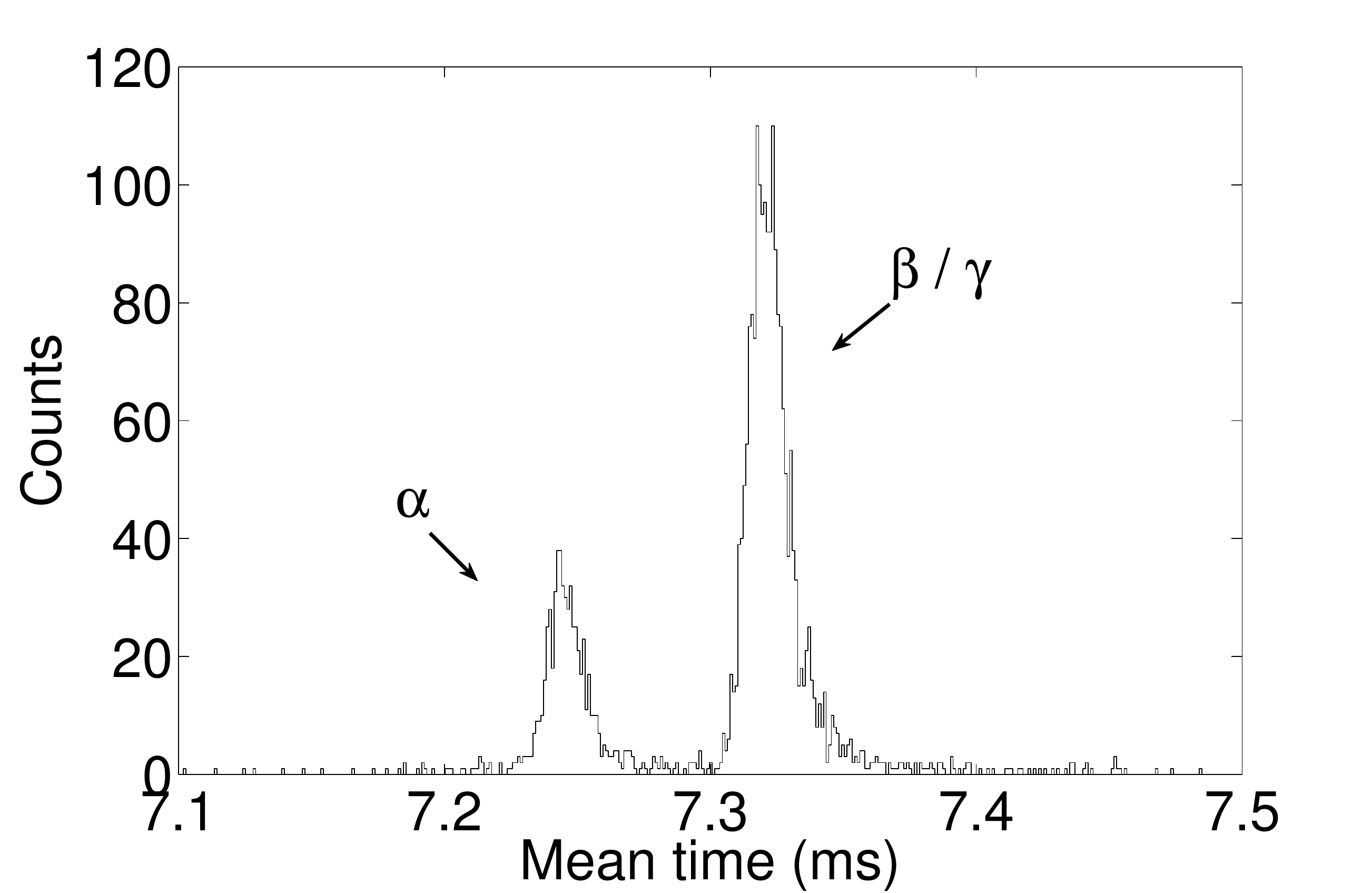}
\caption{Distribution of mean-time parameter in 4 MeV$< E <$ 5 MeV region of alpha-equivalent energy. Discrimination power was found to be 7.6 from fiting each group of distribution with a normal Gaussian function. Right-hand side tail is noticeable for the two groups toward higher mean-time value.}
        \label{fig:PSD_center}
\end{figure}

A two dimensional scatter plot of the pulse heights and mean-times of signals obtained in a 95 h background measurement is shown in Fig.~\ref{fig:PH_meantime}. The pulse height is the difference between the maximum value and baseline level of a signal. The maximum value is found using a quadratic polynomial fit to the region of the signal near the pulse maximum. The baseline level is the average voltage value in the time region before the signal rises. The mean-time parameter is defined as 
\begin{equation}
\centering
t_{\mathrm{mean}}= {{\sum\limits_{t_{10}-l}^{t_{10}+r}({v_t \times (t-t_{10}}})) / {\sum\limits_{t_{10}-l}^{t_{10}+r}v_t}},
\label{eq:meantime}
\end{equation} 
where $v_t$ is the measured voltage value at time $t$ subtracting the baseline level, $t_{10}$ is the time when it reaches 10\% of the pulse height, and $l$ and $r$ indicate the time length of the signal toward left and right directions from $t_{10}$, respectively, to calculate the mean-time. $l$ is set to reach the time at the baseline level, while $r$ is a free parameter that was selected to achieve the most efficient particle discrimination. Here, the $r$ value was set to not include the negative part of an AC coupled signal (see Fig. 2),
which was used for the energy spectrum because it was recorded with finer digitizer resolution (i.e., bigger gain is used) than the DC coupled signal. 
\begin{figure}
        \centering
        \includegraphics[width=1\linewidth,keepaspectratio]{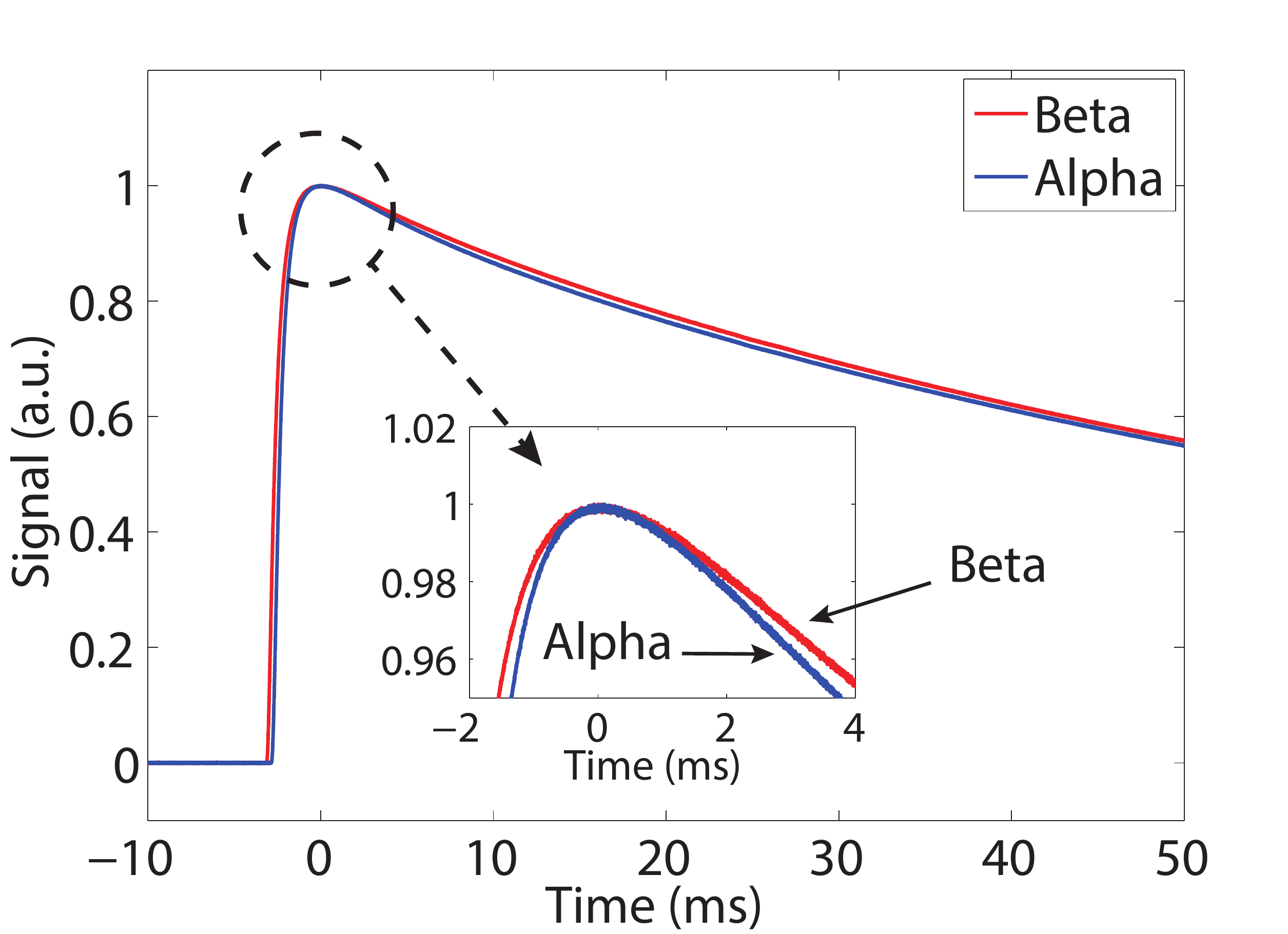}
    \caption{(Color online) Averaged pulse shapes of $\alpha$ induced signals from $^{232}$Th decays (blue) and the corresponding (i.e., with the same pulse height) muon-induced signals (red). The signals are normalized for their pulse height, and shifted in time to align the maximum point of signals.}
        \label{fig:alpha_vs_beta}
\end{figure}

 The pulse-shape discrimination (PSD) between $\alpha$ and $\beta/\gamma$ particles can be realized with the mean-time as a pulse shape parameter due to the difference in the rise and decay times of the two types of events.
A similar tendency of PSD was reported for other low temperature scintillating detectors~\cite{gironi_PSD_app}. 
The separation into two groups of the events in the energy region between 4 and 5~MeV of alpha-equivalent energy can be readily observed from the distribution of mean-time values shown in Fig.~\ref{fig:PSD_center}. 
The energy of $\alpha$ induced events was determined as described in the following section. Although the two peaks have noticeable right-hand tails toward higher mean-time values, normal Gaussian functions were used to fit the distributions. We interpret the right-hand tails as a result of signal pile-up. A parameter of discrimination power ($DP$) is defined as     
\begin{equation}
\centering
DP={({\mu_{1}-\mu_{2}})/\sqrt{\sigma_1^2+\sigma_2^2}},
\end{equation}
where $\mu_i$ are the mean values, $\sigma_i$ are the standard deviations of the Gaussian distributions for $\alpha$ and $\beta/\gamma$ events. DP was found to be 7.6.

The averaged pulse shapes for the two groups of events are compared in Fig.~\ref{fig:alpha_vs_beta}.  
Templates of $\alpha$ events were obtained by averaging out the $^{232}$Th $\alpha$-decay events pulse profiles with energy release in the crystal 4082~keV (due to the contamination of the crystal by thorium), whereas the events caused by cosmic muons with the same pulse height were selected for the template of $\beta$ induced events. 
The normalized pulses of alpha and beta templates are aligned at the time of their maximum values, as shown in Fig.~\ref{fig:alpha_vs_beta}.  
Both the rise and decay times of the $\alpha$ and $\beta$ signals are clearly different.
The signals induced by $\alpha$ events have faster rise and faster decay than those of the $\beta$ events.

 \begin{figure} 
        \centering
                \includegraphics[width=1\linewidth,keepaspectratio]{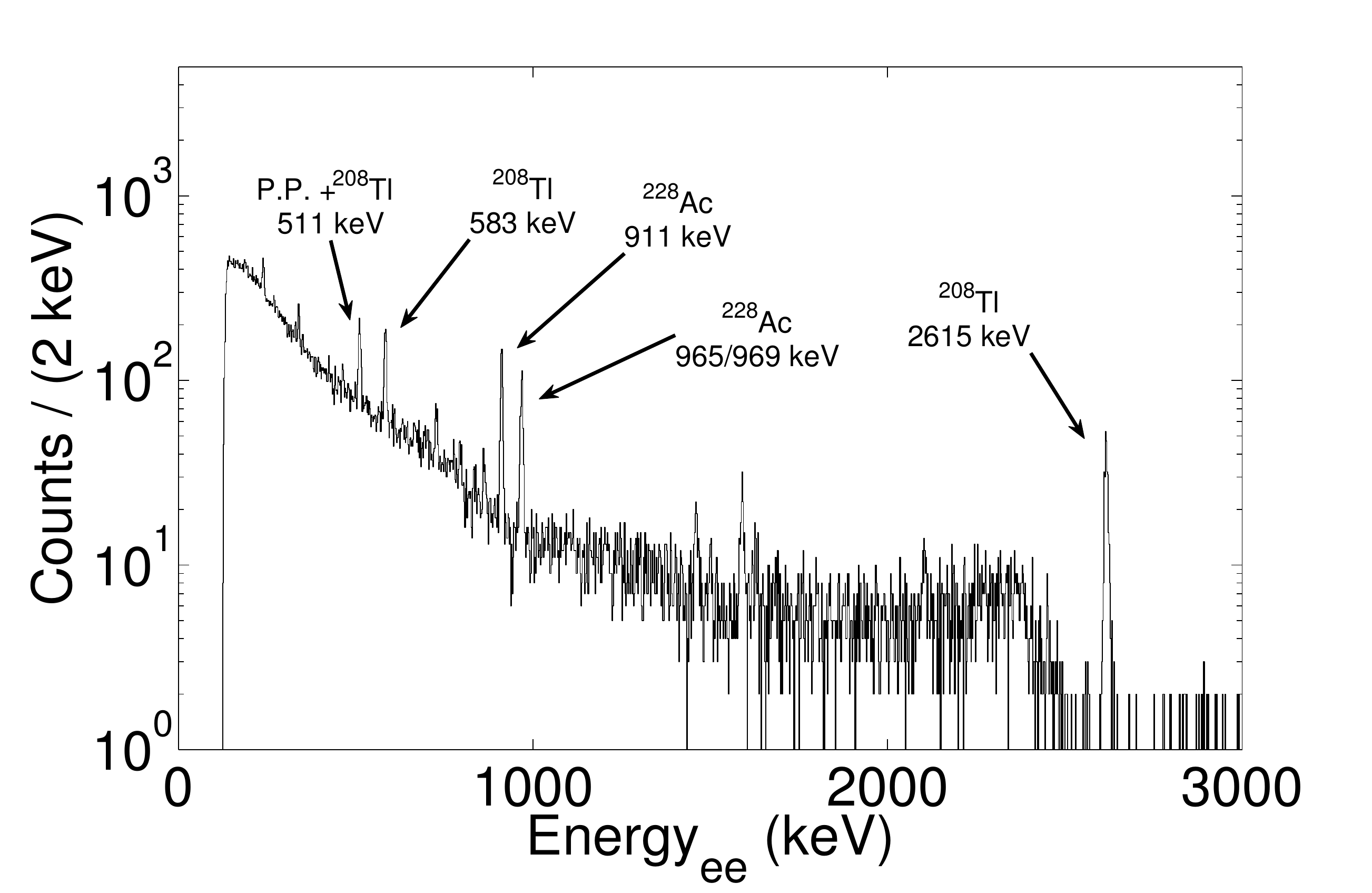}
    \caption{Energy spectrum measured with an external source over 65 h.}
        \label{fig:E_spec}
\end{figure}
According to scintillation measurements of a \CMO{} crystal at 7-300~K~\cite{CMO_decay,Mikhailik_2010}, the scintillation decay-time of the \CMO{} crystal reaches hundreds of $\mu$s at 7 K. The crystal also shows different light output for $\alpha$ and $\beta / \gamma$  events~\cite{Annenkov_CMO_property}. 
A slowly decaying scintillation mechanism would cause slow generation of phonon in the \CMO{} crystal. 
$\alpha$ and $\beta / \gamma$ particle events may have different fractions of slow component for phonon generation. 
This difference in slow phonon generation at mK temperatures may induce different pulse shapes for $\alpha$ and $\beta / \gamma$ particle events.

\section{Energy spectrum}

Because athermal phonon absorption in the phonon collector significantly contributes to the signal size~\cite{gbkim_ltd15}, the signals have some degrees of position dependence for their pulse height and shape. 
In this detector, the signals with faster rise-times show bigger pulse heights for the same energy events. 
This effect can be observed in Fig.~\ref{fig:PH_meantime}. The distribution of mean-time values and pulse heights for alpha and gamma-ray full absorption lines has anti-correlated slopes. This negative slope more dominantly appears in the alpha signals, which is likely because the full-energy $\gamma$ peaks originate from multiple Compton scatterings in the crystal that smear the position dependence on the event location. 
The pulse height is not an optimal parameter to obtain a high resolution. 
In the optimal filtering method~\cite{yuryev_nim} that is often adopted in high resolution micro-calorimeters,  the signals are assumed to have one shape but with different amplitudes.  
Thus, the optimal filtering method is not applicable to provide a high resolution spectrum for the present signals. 
\begin{figure} 
        \centering
                \includegraphics[width=1\linewidth,keepaspectratio]{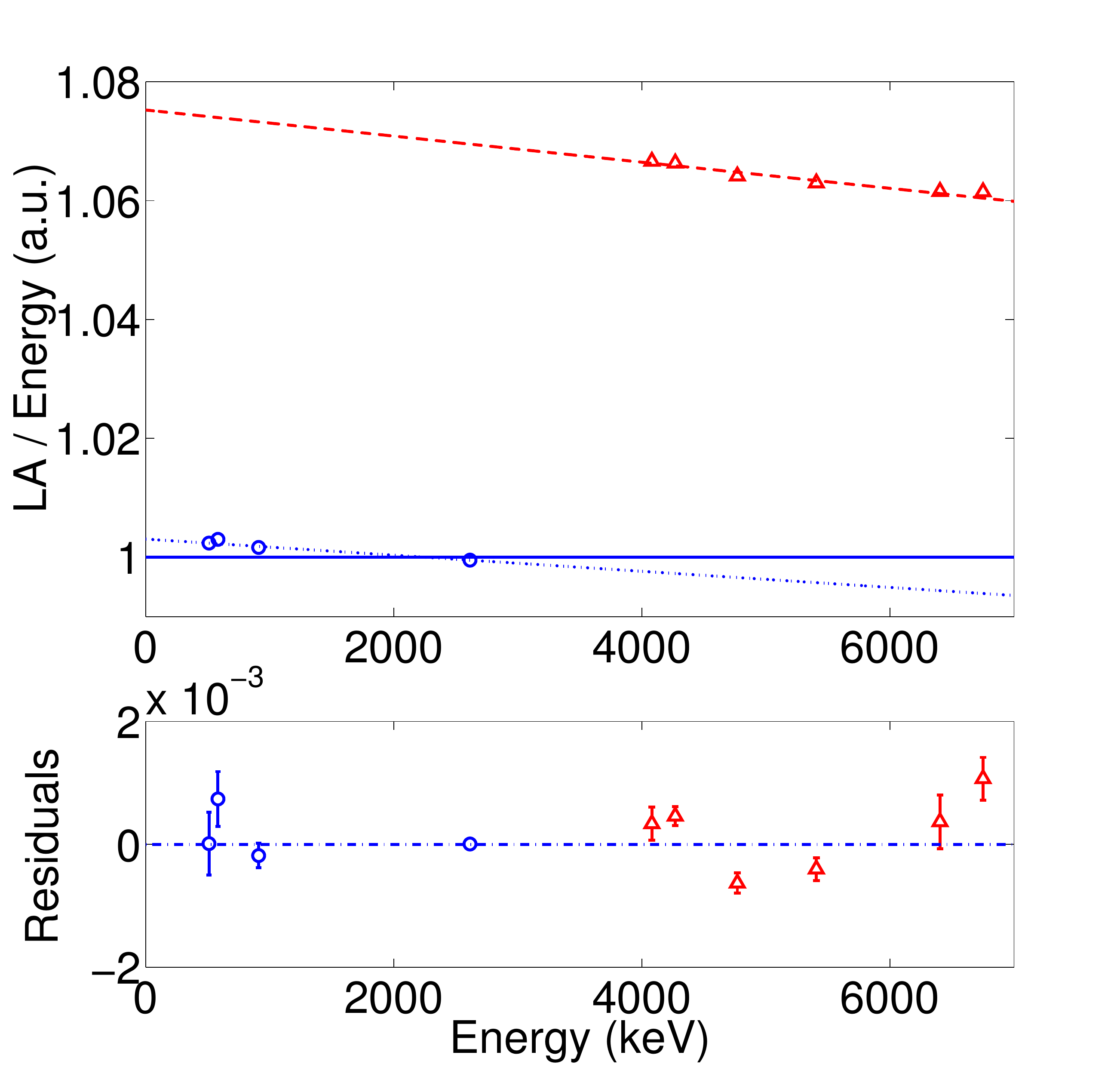}
    \caption{(Color online) Ratio of signal amplitude ($LA$) and energy in linear calibration for $\gamma$ (circle, blue) and $\alpha$ (triangle, red) peaks (upper). Quadratic calibration lines (dotted  and dashed lines) are shown for $\gamma$ and $\alpha$ peaks, respectively. Deviations of the ratios from each quadratic functions are shown in the lower figure.}
        \label{fig:cal_scale}
\end{figure}

In the present analysis, a new parameter, $LA$ (left area) was used as an amplitude parameter to reduce the position dependence effect of the large crystal detector. It is defined as
\begin{equation}
\centering
LA = {\sum\limits_{t_{10}-l}^{t_{\mathrm{mean}}}{v_t}
}
\end{equation}
where the variables are as defined in Eq.~\ref{eq:meantime}.
$LA$ is a partial integration for the leading part of a signal. 
Because the integration range of the $LA$ parameter is set by a shape-dependent parameter, the mean-time, this parameter is less influenced by the pulse shape. 
For instance, the correlation coefficient between the pulse height and mean-time was $-$0.62 for 4082~keV $\alpha$ signals, but it was 0.08 between the $LA$ and mean-time.

In a calibration run, a thoriated tungsten rod was used as an external gamma-ray source. The source was placed in the gap between the cryostat and the lead shield.

When a linear energy calibration was applied to gamma-ray peaks, deviations from the linear calibration of less than 0.4\% were found for low-energy peaks. 
A quadratic function with no constant term was used for the calibration of electron-equivalent energy for 511, 583, 911 and 2615 keV peaks for the spectrum shown in Fig. 6.
The corresponding energy resolutions of the peaks are listed in Table~\ref{tab:FWHM}. 

Fig.~\ref{fig:cal_scale} shows the linearities of the electron and alpha signals. The $LA$ values of the electron and alpha peaks divided by the linear calibration of the gamma-ray peaks are plotted in the upper figure. The $LA$/energy ratios for the alpha peaks are about 6\% larger than those for the gamma-ray peaks. The quadratic fit functions are shown as dotted and dashed lines for gamma and alpha peaks, respectively. The residuals in the lower figure indicate deviations from the quadratic functions for the two groups. 
With the quadratic calibration, a very small deviation is expected near 3~MeV for electron measurements.
It implies that this method can supply an accurate energy calibration at the $Q_{\beta\beta}$ value of the \twodbd{} of $^{100}$Mo.

\begin{figure}
        \centering
                \includegraphics[width=1\linewidth,keepaspectratio]{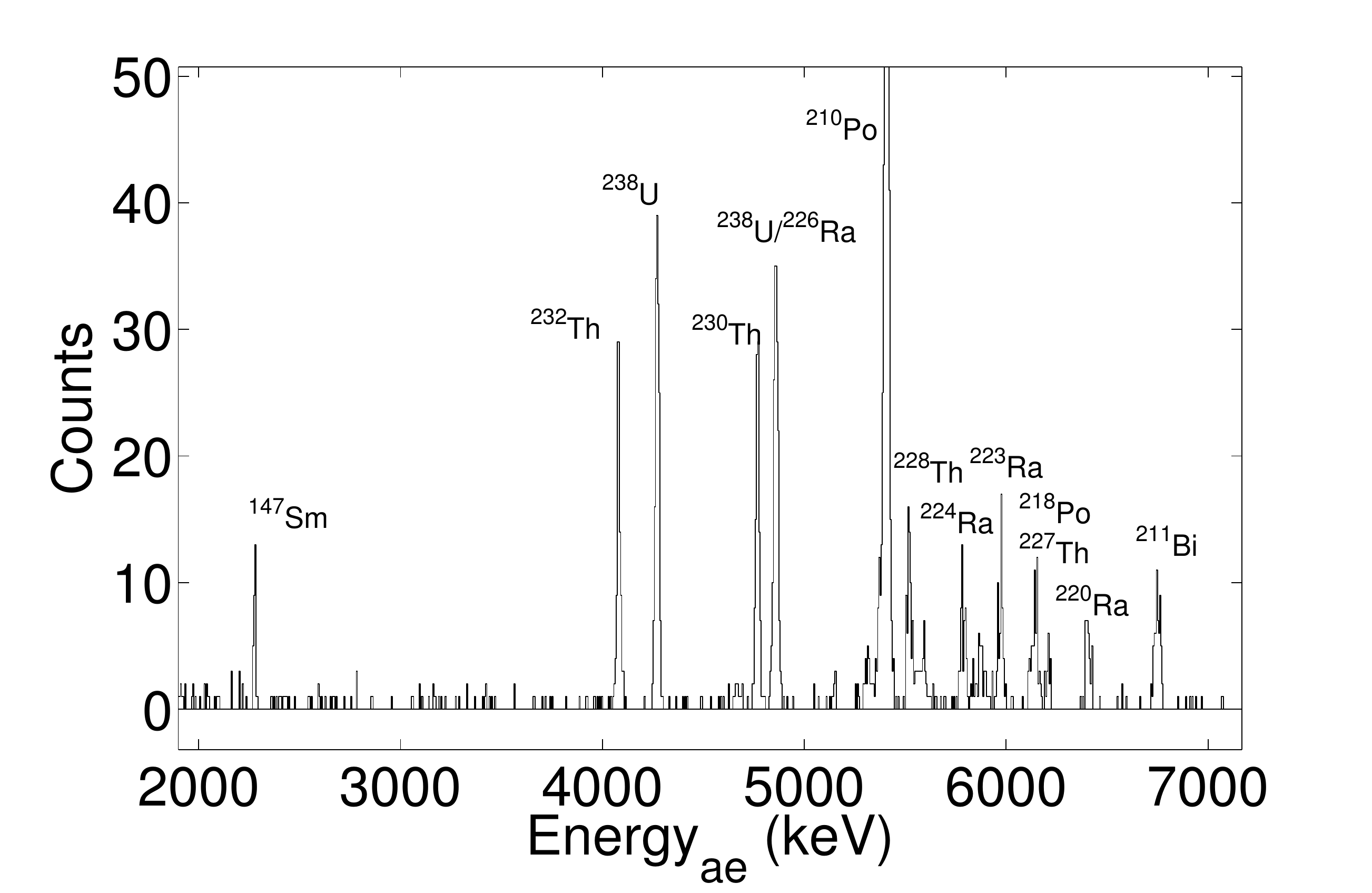}
    \caption{Energy spectrum of bulk alpha events in a background measurement.}
        \label{fig:alpha_spec}
\end{figure}
\begin{table} [b]
\caption{Energy resolutions of gamma-ray full absorption peaks in the calibration measurement.}
\centering
\begin{tabular}{c|c}
\hline\hline
Energy (keV) & FWHM (keV)  \\ [0.5ex]
\hline
511 & 7.2 $\pm$ 0.6 \\
583 & 6.5 $\pm$ 0.5 \\
911 & 6.8 $\pm$ 0.3 \\
2615 & 10.9 $\pm$ 0.4 \\
\hline
\end{tabular}
\label{tab:FWHM}
\end{table}

Alpha events can be separated from $\beta/\gamma$ events using the mean-time parameter. The energy spectrum of alpha events is shown in Fig.~\ref{fig:alpha_spec}. 
An energy calibration for this spectrum was performed with a quadratic function for the alpha peaks as discussed above.
These background were bulk events of alpha decays in the crystal, mainly from radionuclides of U and Th decay chains, most of them are identified as shown in Fig.~\ref{fig:alpha_spec}.
Because this crystal was developed to investigate the scintillation properties of \CMO{}, its internal background was not exceptionally low. The AMoRE collaboration has developed  \enCMO{} growing technology with low radio-impurities. Internal alpha activities of about 80~$\mu$Bq/kg of $^{226}$Ra and 70~$\mu$Bq/kg of $^{228}$Th were found in a 196~g \enCMO~\cite{jhso_CMO_background}. 

Not only alpha signals can be rejected efficiently, but also $\beta$ decays of $^{212}$Bi, $^{214}$Bi and $^{208}$Tl can be tagged and eliminated from the data by using information about associated alpha-emitting nuclides. Random pile-ups of events (first of all from the \twodbd) could be a substantial source of background of LTD to search for \dbd{} due to the poor time resolution~\cite{chernyak_random_coin_dbd}. Relatively fast response of the MMC among the LTDs is a certain advantage to discriminate the background.

\section{Conclusion}

LTDs made of crystal scintillators containing isotopes of interest have distinct advantages in \dbd{} searches. 
Such LTDs make it possible to provide a high detection efficiency to the \dbd. Taking the advantage of high resolution sensor technologies, these dielectric detectors in the heat (phonon) measurement can have comparable energy resolution to those of HPGe detectors.  
The comparison in heat/light measurement channels makes unambiguous separation of alpha events from electron events. 
As discussed in this report, pulse shape discrimination is also possible using only the phonon measurement. This PSD capability of the phonon sensor increases discrimination power for alpha background signals, or can simplify the detector cell design by using only one phonon sensor without a photon sensor that is commonly used for particle discrimination~\cite{artusa2014exploring}, and will reduce the number of measurement channels. 

In comparison with signals from PMTs or conventional semiconductor detectors, phonon signals from a crystal detector in the LTD concept are typically slow. Even though the energy resolution of these detectors does not suffer from the slow signal in the low activity environment of an underground experiment, random coincidence of events lead to an unavoidable background because of the slow rise-time~\cite{chernyak_random_coin_dbd}. Two consecutive electron events that occur in a time interval that is much shorter than the signal rise-time can be regarded as a single event. Such randomly coincident events, particularly of \twodbd, are an unavoidable source of backgrounds in the \dbd{} taking into account that the \twodbd{} event rate of $^{100}$Mo is expected to be about 10~mBq in 1~kg \enCMO.   

In the present experiment, the phonon signals had rise-times of 1.1~ms, which is much faster than the rise-times of LTDs with NTD Ge thermistors. The fast rise-time provides efficient rejection possibility for randomly coincident events~\cite{chernyak_rej_random_coin}. Moreover, a photon detector composed of a 2 inch Ge wafer and an MMC sensor showed a temperature independent rise-time of about 0.2 ms with reasonable energy resolution \cite{hjlee2015}. Simultaneous measurements with the photon detector will increase  the discrimination power not just for alpha events but also for randomly coincident events.  
 
For the AMoRE project, \enCMO{} crystals will be used as the detector material together with MMCs in phonon and photon measurement setups. We aim to reach a zero background with improved energy resolution of a few keV.
The first stage experiment is expected to be constructed with a 10 kg prototype detector by 2016. We plan to perform a large scale experiment with 200 kg \enCMO{} crystals in the next 5-6 years.
The sensitivity of the experiment to the effective Majorana neutrino mass is estimated to be in the range of 20-50 meV, which corresponds to the inverted scheme of the neutrino mass. 

\section*{Conflict of Interests}
The authors declare that there is no conflict of interests regarding the publication of this paper.

\section*{Acknowledgment}
This research was funded by Grant No. IBS-R016-G1, and partly supported by the National Research Foundation of Korea Grant funded by the Korean Government (NRF-2011-220-C00006, and NRF-2013K2A5A3000039). 
F.A. Danevich was supported in part by the Space Research Program of the National Academy of Sciences of Ukraine.

% \bibliography{gbkim_cmo_psd}

\begin{thebibliography}{32}
\expandafter\ifx\csname natexlab\endcsname\relax\def\natexlab#1{#1}\fi
\expandafter\ifx\csname bibnamefont\endcsname\relax
  \def\bibnamefont#1{#1}\fi
\expandafter\ifx\csname bibfnamefont\endcsname\relax
  \def\bibfnamefont#1{#1}\fi
\expandafter\ifx\csname citenamefont\endcsname\relax
  \def\citenamefont#1{#1}\fi
\expandafter\ifx\csname url\endcsname\relax
  \def\url#1{\texttt{#1}}\fi
\expandafter\ifx\csname urlprefix\endcsname\relax\def\urlprefix{URL }\fi
\providecommand{\bibinfo}[2]{#2}
\providecommand{\eprint}[2][]{\url{#2}}

\bibitem[{\citenamefont{Beringer et~al.}(2012)\citenamefont{Beringer, Arguin,
  Barnett, Copic, Dahl, Groom, Lin, Lys, Murayama, Wohl et~al.}}]{rev_pp}
\bibinfo{author}{\bibfnamefont{J.}~\bibnamefont{Beringer}},
  \bibinfo{author}{\bibfnamefont{J.~F.} \bibnamefont{Arguin}},
  \bibinfo{author}{\bibfnamefont{R.~M.} \bibnamefont{Barnett}},
  \bibinfo{author}{\bibfnamefont{K.}~\bibnamefont{Copic}},
  \bibinfo{author}{\bibfnamefont{O.}~\bibnamefont{Dahl}},
  \bibinfo{author}{\bibfnamefont{D.~E.} \bibnamefont{Groom}},
  \bibinfo{author}{\bibfnamefont{C.~J.} \bibnamefont{Lin}},
  \bibinfo{author}{\bibfnamefont{J.}~\bibnamefont{Lys}},
  \bibinfo{author}{\bibfnamefont{H.}~\bibnamefont{Murayama}},
  \bibinfo{author}{\bibfnamefont{C.~G.} \bibnamefont{Wohl}},
  \bibnamefont{et~al.} (\bibinfo{collaboration}{Particle Data Group}),
  \bibinfo{journal}{Phys. Rev. D} \textbf{\bibinfo{volume}{86}},
  \bibinfo{pages}{010001} (\bibinfo{year}{2012}).

\bibitem[{\citenamefont{Mohapatra et~al.}(2007)\citenamefont{Mohapatra,
  Antusch, Babu, Barenboim, Chen, de~Gouv{\^{e}}a, de~Holanda, Dutta, Grossman,
  Joshipura et~al.}}]{Mohapatra_theory_neu}
\bibinfo{author}{\bibfnamefont{R.~N.} \bibnamefont{Mohapatra}},
  \bibinfo{author}{\bibfnamefont{S.}~\bibnamefont{Antusch}},
  \bibinfo{author}{\bibfnamefont{K.~S.} \bibnamefont{Babu}},
  \bibinfo{author}{\bibfnamefont{G.}~\bibnamefont{Barenboim}},
  \bibinfo{author}{\bibfnamefont{M.-C.} \bibnamefont{Chen}},
  \bibinfo{author}{\bibfnamefont{A.}~\bibnamefont{de~Gouv{\^{e}}a}},
  \bibinfo{author}{\bibfnamefont{P.}~\bibnamefont{de~Holanda}},
  \bibinfo{author}{\bibfnamefont{B.}~\bibnamefont{Dutta}},
  \bibinfo{author}{\bibfnamefont{Y.}~\bibnamefont{Grossman}},
  \bibinfo{author}{\bibfnamefont{A.}~\bibnamefont{Joshipura}},
  \bibnamefont{et~al.}, \bibinfo{journal}{Rep. Prog. Phys.}
  \textbf{\bibinfo{volume}{70}}, \bibinfo{pages}{1757} (\bibinfo{year}{2007}).

\bibitem[{\citenamefont{Elliott and Vogel}(2002)}]{elliott_dbd}
\bibinfo{author}{\bibfnamefont{S.~R.} \bibnamefont{Elliott}} \bibnamefont{and}
  \bibinfo{author}{\bibfnamefont{P.}~\bibnamefont{Vogel}},
  \bibinfo{journal}{Annu. Rev. Nucl. Part. Sci.} \textbf{\bibinfo{volume}{52}},
  \bibinfo{pages}{115} (\bibinfo{year}{2002}).

\bibitem[{\citenamefont{Avignone et~al.}(2008)\citenamefont{Avignone, Elliott,
  and Engel}}]{Avignone_dbd}
\bibinfo{author}{\bibfnamefont{F.~T.} \bibnamefont{Avignone}},
  \bibinfo{author}{\bibfnamefont{S.~R.} \bibnamefont{Elliott}},
  \bibnamefont{and} \bibinfo{author}{\bibfnamefont{J.}~\bibnamefont{Engel}},
  \bibinfo{journal}{Rev. Mod. Phys.} \textbf{\bibinfo{volume}{80}},
  \bibinfo{pages}{481} (\bibinfo{year}{2008}).

\bibitem[{\citenamefont{Rodejohann}(2011)}]{rodejohann_dbd}
\bibinfo{author}{\bibfnamefont{W.}~\bibnamefont{Rodejohann}},
  \bibinfo{journal}{Int. J. Mod. Phys. A} \textbf{\bibinfo{volume}{20}},
  \bibinfo{pages}{1833} (\bibinfo{year}{2011}).

\bibitem[{\citenamefont{Giuliani and Poves}(2012)}]{andrea_ahep}
\bibinfo{author}{\bibfnamefont{A.}~\bibnamefont{Giuliani}} \bibnamefont{and}
  \bibinfo{author}{\bibfnamefont{A.}~\bibnamefont{Poves}},
  \bibinfo{journal}{Adv. High Energy Phys.} \textbf{\bibinfo{volume}{2012}}
  (\bibinfo{year}{2012}).

\bibitem[{\citenamefont{G{\'{o}}mez-Cadenas
  et~al.}(2012)\citenamefont{G{\'{o}}mez-Cadenas, Mart{\'{i}}n-Albo, Mezzetto,
  Monrabal, and Sorel}}]{gomez2012search}
\bibinfo{author}{\bibfnamefont{J.}~\bibnamefont{G{\'{o}}mez-Cadenas}},
  \bibinfo{author}{\bibfnamefont{J.}~\bibnamefont{Mart{\'{i}}n-Albo}},
  \bibinfo{author}{\bibfnamefont{M.}~\bibnamefont{Mezzetto}},
  \bibinfo{author}{\bibfnamefont{F.}~\bibnamefont{Monrabal}}, \bibnamefont{and}
  \bibinfo{author}{\bibfnamefont{M.}~\bibnamefont{Sorel}},
  \bibinfo{journal}{RIVISTA DEL NUOVO CIMENTO} \textbf{\bibinfo{volume}{35}},
  \bibinfo{pages}{29} (\bibinfo{year}{2012}).

\bibitem[{\citenamefont{Enss}(2005)}]{enss2005cryogenic}
\bibinfo{author}{\bibfnamefont{C.}~\bibnamefont{Enss}},
  \emph{\bibinfo{title}{Cryogenic particle detection}},
  vol.~\bibinfo{volume}{99} (\bibinfo{publisher}{Springer},
  \bibinfo{year}{2005}).

\bibitem[{\citenamefont{Bhang et~al.}(2012)\citenamefont{Bhang, Boiko,
  Chernyak, Choi, Choi, Danevich, Efendiev, Enss, Fleischmann, Gangapshev
  et~al.}}]{bhang_AMoRE}
\bibinfo{author}{\bibfnamefont{H.}~\bibnamefont{Bhang}},
  \bibinfo{author}{\bibfnamefont{R.~S.} \bibnamefont{Boiko}},
  \bibinfo{author}{\bibfnamefont{D.~M.} \bibnamefont{Chernyak}},
  \bibinfo{author}{\bibfnamefont{J.~H.} \bibnamefont{Choi}},
  \bibinfo{author}{\bibfnamefont{S.}~\bibnamefont{Choi}},
  \bibinfo{author}{\bibfnamefont{F.~A.} \bibnamefont{Danevich}},
  \bibinfo{author}{\bibfnamefont{K.~V.} \bibnamefont{Efendiev}},
  \bibinfo{author}{\bibfnamefont{C.}~\bibnamefont{Enss}},
  \bibinfo{author}{\bibfnamefont{A.}~\bibnamefont{Fleischmann}},
  \bibinfo{author}{\bibfnamefont{A.~M.} \bibnamefont{Gangapshev}},
  \bibnamefont{et~al.}, \bibinfo{journal}{J. Phys. Conf. Ser.}
  \textbf{\bibinfo{volume}{375}}, \bibinfo{pages}{042023}
  (\bibinfo{year}{2012}).

\bibitem[{\citenamefont{Lee et~al.}(2011)\citenamefont{Lee, Choi, Danevich,
  Jang, Kang, Khanbekov, Kim, Kim, Kim, Kim et~al.}}]{sjlee_app}
\bibinfo{author}{\bibfnamefont{S.~J.} \bibnamefont{Lee}},
  \bibinfo{author}{\bibfnamefont{J.~H.} \bibnamefont{Choi}},
  \bibinfo{author}{\bibfnamefont{F.~A.} \bibnamefont{Danevich}},
  \bibinfo{author}{\bibfnamefont{Y.~S.} \bibnamefont{Jang}},
  \bibinfo{author}{\bibfnamefont{W.~G.} \bibnamefont{Kang}},
  \bibinfo{author}{\bibfnamefont{N.}~\bibnamefont{Khanbekov}},
  \bibinfo{author}{\bibfnamefont{H.~J.} \bibnamefont{Kim}},
  \bibinfo{author}{\bibfnamefont{I.~H.} \bibnamefont{Kim}},
  \bibinfo{author}{\bibfnamefont{S.~C.} \bibnamefont{Kim}},
  \bibinfo{author}{\bibfnamefont{S.~K.} \bibnamefont{Kim}},
  \bibnamefont{et~al.}, \bibinfo{journal}{Astropart. Phys.}
  \textbf{\bibinfo{volume}{34}}, \bibinfo{pages}{732 } (\bibinfo{year}{2011}).

\bibitem[{\citenamefont{Kim et~al.}(2014)\citenamefont{Kim, Choi, Jang, Kim,
  Kim, Kobychev, Lee, Lee, Lee, Lee et~al.}}]{gbkim_ltd15}
\bibinfo{author}{\bibfnamefont{G.~B.} \bibnamefont{Kim}},
  \bibinfo{author}{\bibfnamefont{S.}~\bibnamefont{Choi}},
  \bibinfo{author}{\bibfnamefont{Y.~S.} \bibnamefont{Jang}},
  \bibinfo{author}{\bibfnamefont{H.~J.} \bibnamefont{Kim}},
  \bibinfo{author}{\bibfnamefont{Y.~H.} \bibnamefont{Kim}},
  \bibinfo{author}{\bibfnamefont{V.~V.} \bibnamefont{Kobychev}},
  \bibinfo{author}{\bibfnamefont{H.~J.} \bibnamefont{Lee}},
  \bibinfo{author}{\bibfnamefont{J.~H.} \bibnamefont{Lee}},
  \bibinfo{author}{\bibfnamefont{J.~Y.} \bibnamefont{Lee}},
  \bibinfo{author}{\bibfnamefont{M.~K.} \bibnamefont{Lee}},
  \bibnamefont{et~al.}, \bibinfo{journal}{J. Low Temp. Phys.}
  \textbf{\bibinfo{volume}{176}}, \bibinfo{pages}{637} (\bibinfo{year}{2014}).

\bibitem[{\citenamefont{Kim et~al.}(2010)\citenamefont{Kim, Annenkov, Boiko,
  Buzanov, Chernyak, Cho, Danevich, Dossovitsky, Rooh, Kang
  et~al.}}]{hjkim_IEEE_2010}
\bibinfo{author}{\bibfnamefont{H.~J.} \bibnamefont{Kim}},
  \bibinfo{author}{\bibfnamefont{A.~N.} \bibnamefont{Annenkov}},
  \bibinfo{author}{\bibfnamefont{R.~S.} \bibnamefont{Boiko}},
  \bibinfo{author}{\bibfnamefont{O.~A.} \bibnamefont{Buzanov}},
  \bibinfo{author}{\bibfnamefont{D.~M.} \bibnamefont{Chernyak}},
  \bibinfo{author}{\bibfnamefont{J.~H.} \bibnamefont{Cho}},
  \bibinfo{author}{\bibfnamefont{F.~A.} \bibnamefont{Danevich}},
  \bibinfo{author}{\bibfnamefont{A.~E.} \bibnamefont{Dossovitsky}},
  \bibinfo{author}{\bibfnamefont{G.}~\bibnamefont{Rooh}},
  \bibinfo{author}{\bibfnamefont{U.~K.} \bibnamefont{Kang}},
  \bibnamefont{et~al.}, \bibinfo{journal}{IEEE trans. Nucl. Sci.}
  \textbf{\bibinfo{volume}{57}}, \bibinfo{pages}{1475} (\bibinfo{year}{2010}).

\bibitem[{\citenamefont{Pirro et~al.}(2006)\citenamefont{Pirro, Beeman,
  Capelli, Pavan, Previtali, and Gorla}}]{pirro_2006}
\bibinfo{author}{\bibfnamefont{S.}~\bibnamefont{Pirro}},
  \bibinfo{author}{\bibfnamefont{J.}~\bibnamefont{Beeman}},
  \bibinfo{author}{\bibfnamefont{S.}~\bibnamefont{Capelli}},
  \bibinfo{author}{\bibfnamefont{M.}~\bibnamefont{Pavan}},
  \bibinfo{author}{\bibfnamefont{E.}~\bibnamefont{Previtali}},
  \bibnamefont{and} \bibinfo{author}{\bibfnamefont{P.}~\bibnamefont{Gorla}},
  \bibinfo{journal}{Phys. At. Nucl.} \textbf{\bibinfo{volume}{69}},
  \bibinfo{pages}{2109} (\bibinfo{year}{2006}).

\bibitem[{\citenamefont{Rahaman et~al.}(2008)\citenamefont{Rahaman, Elomaa,
  Eronen, Hakala, Jokinen, Julin, Kankainen, Saastamoinen, Suhonen, Weber
  et~al.}}]{Rahaman2008111}
\bibinfo{author}{\bibfnamefont{S.}~\bibnamefont{Rahaman}},
  \bibinfo{author}{\bibfnamefont{V.-V.} \bibnamefont{Elomaa}},
  \bibinfo{author}{\bibfnamefont{T.}~\bibnamefont{Eronen}},
  \bibinfo{author}{\bibfnamefont{J.}~\bibnamefont{Hakala}},
  \bibinfo{author}{\bibfnamefont{A.}~\bibnamefont{Jokinen}},
  \bibinfo{author}{\bibfnamefont{J.}~\bibnamefont{Julin}},
  \bibinfo{author}{\bibfnamefont{A.}~\bibnamefont{Kankainen}},
  \bibinfo{author}{\bibfnamefont{A.}~\bibnamefont{Saastamoinen}},
  \bibinfo{author}{\bibfnamefont{J.}~\bibnamefont{Suhonen}},
  \bibinfo{author}{\bibfnamefont{C.}~\bibnamefont{Weber}},
  \bibnamefont{et~al.}, \bibinfo{journal}{Phys. Lett. B}
  \textbf{\bibinfo{volume}{662}}, \bibinfo{pages}{111 } (\bibinfo{year}{2008}).

\bibitem[{\citenamefont{Wieser and Laeter}(2007)}]{Mo_abundance}
\bibinfo{author}{\bibfnamefont{M.~E.} \bibnamefont{Wieser}} \bibnamefont{and}
  \bibinfo{author}{\bibfnamefont{J.~R.~D.} \bibnamefont{Laeter}},
  \bibinfo{journal}{Phys. Rev. C} \textbf{\bibinfo{volume}{75}},
  \bibinfo{pages}{055802} (\bibinfo{year}{2007}).

\bibitem[{\citenamefont{Barea et~al.}(2012)\citenamefont{Barea, Kotila, and
  Iachello}}]{barea2012_prl_109_042501_2012}
\bibinfo{author}{\bibfnamefont{J.}~\bibnamefont{Barea}},
  \bibinfo{author}{\bibfnamefont{J.}~\bibnamefont{Kotila}}, \bibnamefont{and}
  \bibinfo{author}{\bibfnamefont{F.}~\bibnamefont{Iachello}},
  \bibinfo{journal}{Phys. Rev. Lett.} \textbf{\bibinfo{volume}{109}},
  \bibinfo{pages}{042501} (\bibinfo{year}{2012}).

\bibitem[{\citenamefont{Vergados et~al.}(2012)\citenamefont{Vergados, Ejiri,
  and F{\v{s}}imkovic}}]{Vergados_2012}
\bibinfo{author}{\bibfnamefont{J.~D.} \bibnamefont{Vergados}},
  \bibinfo{author}{\bibfnamefont{H.}~\bibnamefont{Ejiri}}, \bibnamefont{and}
  \bibinfo{author}{\bibnamefont{F{\v{s}}imkovic}}, \bibinfo{journal}{Rep. Prog.
  Phys.} \textbf{\bibinfo{volume}{75}}, \bibinfo{pages}{106301}
  (\bibinfo{year}{2012}).

\bibitem[{\citenamefont{So et~al.}(2012)\citenamefont{So, Kim, Alenkov,
  Annenkov, Bhang, Boiko, Buzanov, Chernyak, Choi, Choi
  et~al.}}]{jhso_CMO_background}
\bibinfo{author}{\bibfnamefont{J.~H.} \bibnamefont{So}},
  \bibinfo{author}{\bibfnamefont{H.~J.} \bibnamefont{Kim}},
  \bibinfo{author}{\bibfnamefont{V.~V.} \bibnamefont{Alenkov}},
  \bibinfo{author}{\bibfnamefont{A.~N.} \bibnamefont{Annenkov}},
  \bibinfo{author}{\bibfnamefont{H.}~\bibnamefont{Bhang}},
  \bibinfo{author}{\bibfnamefont{R.~S.} \bibnamefont{Boiko}},
  \bibinfo{author}{\bibfnamefont{O.~A.} \bibnamefont{Buzanov}},
  \bibinfo{author}{\bibfnamefont{D.~M.} \bibnamefont{Chernyak}},
  \bibinfo{author}{\bibfnamefont{J.~H.} \bibnamefont{Choi}},
  \bibinfo{author}{\bibfnamefont{S.}~\bibnamefont{Choi}}, \bibnamefont{et~al.},
  \bibinfo{journal}{IEEE trans. Nucl. Sci.} \textbf{\bibinfo{volume}{59}},
  \bibinfo{pages}{2214} (\bibinfo{year}{2012}).

\bibitem[{\citenamefont{Yoon et~al.}(2012)\citenamefont{Yoon, Jang, Kim, Kim,
  Kim, Lee, Lee, Lee, Lee, Lee et~al.}}]{wsyoon_ltd14}
\bibinfo{author}{\bibfnamefont{W.~S.} \bibnamefont{Yoon}},
  \bibinfo{author}{\bibfnamefont{Y.~S.} \bibnamefont{Jang}},
  \bibinfo{author}{\bibfnamefont{G.~B.} \bibnamefont{Kim}},
  \bibinfo{author}{\bibfnamefont{K.~J.} \bibnamefont{Kim}},
  \bibinfo{author}{\bibfnamefont{M.~S.} \bibnamefont{Kim}},
  \bibinfo{author}{\bibfnamefont{J.~S.} \bibnamefont{Lee}},
  \bibinfo{author}{\bibfnamefont{K.~B.} \bibnamefont{Lee}},
  \bibinfo{author}{\bibfnamefont{M.~K.} \bibnamefont{Lee}},
  \bibinfo{author}{\bibfnamefont{S.~J.} \bibnamefont{Lee}},
  \bibinfo{author}{\bibfnamefont{H.~J.} \bibnamefont{Lee}},
  \bibnamefont{et~al.}, \bibinfo{journal}{J. Low Temp. Phys.}
  \textbf{\bibinfo{volume}{167}}, \bibinfo{pages}{280} (\bibinfo{year}{2012}).

\bibitem[{\citenamefont{Yoon et~al.}(2014)\citenamefont{Yoon, Kim, Lee, Lee,
  Lee, Jang, Lee, Lee, and Kim}}]{wsyoon_ltd15}
\bibinfo{author}{\bibfnamefont{W.~S.} \bibnamefont{Yoon}},
  \bibinfo{author}{\bibfnamefont{G.~B.} \bibnamefont{Kim}},
  \bibinfo{author}{\bibfnamefont{H.~J.} \bibnamefont{Lee}},
  \bibinfo{author}{\bibfnamefont{J.~Y.} \bibnamefont{Lee}},
  \bibinfo{author}{\bibfnamefont{J.~H.} \bibnamefont{Lee}},
  \bibinfo{author}{\bibfnamefont{Y.~S.} \bibnamefont{Jang}},
  \bibinfo{author}{\bibfnamefont{S.~J.} \bibnamefont{Lee}},
  \bibinfo{author}{\bibfnamefont{M.~K.} \bibnamefont{Lee}}, \bibnamefont{and}
  \bibinfo{author}{\bibfnamefont{Y.~H.} \bibnamefont{Kim}},
  \bibinfo{journal}{J. Low Temp. Phys.} \textbf{\bibinfo{volume}{176}},
  \bibinfo{pages}{644} (\bibinfo{year}{2014}).

\bibitem[{\citenamefont{Wolfe}(2005)}]{wolfe}
\bibinfo{author}{\bibfnamefont{J.~P.} \bibnamefont{Wolfe}},
  \emph{\bibinfo{title}{Imaging phonons: acoustic wave propagation in solids}}
  (\bibinfo{publisher}{Cambridge University Press}, \bibinfo{year}{2005}).

\bibitem[{\citenamefont{Leman}(2012)}]{leman}
\bibinfo{author}{\bibfnamefont{S.~W.} \bibnamefont{Leman}},
  \bibinfo{journal}{Rev. Sci. Instr.} \textbf{\bibinfo{volume}{83}},
  \bibinfo{pages}{091101} (\bibinfo{year}{2012}).

\bibitem[{\citenamefont{Kim et~al.}(2004)\citenamefont{Kim, Eguchi, Enss,
  Huang, Lanou, Maris, Mocharnuk-Macchia, Seidel, Sethumadhavan, and
  Yao}}]{yhkim2004}
\bibinfo{author}{\bibfnamefont{Y.~H.} \bibnamefont{Kim}},
  \bibinfo{author}{\bibfnamefont{H.}~\bibnamefont{Eguchi}},
  \bibinfo{author}{\bibfnamefont{C.}~\bibnamefont{Enss}},
  \bibinfo{author}{\bibfnamefont{Y.~H.} \bibnamefont{Huang}},
  \bibinfo{author}{\bibfnamefont{R.~E.} \bibnamefont{Lanou}},
  \bibinfo{author}{\bibfnamefont{H.~J.} \bibnamefont{Maris}},
  \bibinfo{author}{\bibfnamefont{A.~N.} \bibnamefont{Mocharnuk-Macchia}},
  \bibinfo{author}{\bibfnamefont{G.~M.} \bibnamefont{Seidel}},
  \bibinfo{author}{\bibfnamefont{B.}~\bibnamefont{Sethumadhavan}},
  \bibnamefont{and} \bibinfo{author}{\bibfnamefont{W.}~\bibnamefont{Yao}},
  \bibinfo{journal}{Nucl. Instr. and Mech. A} \textbf{\bibinfo{volume}{520}},
  \bibinfo{pages}{208} (\bibinfo{year}{2004}).

\bibitem[{\citenamefont{Arnaboldi et~al.}(2011)\citenamefont{Arnaboldi,
  Brofferio, Cremonesi, Gironi, Pavan, Pessina, Pirro, and
  Previtali}}]{gironi_PSD_app}
\bibinfo{author}{\bibfnamefont{C.}~\bibnamefont{Arnaboldi}},
  \bibinfo{author}{\bibfnamefont{C.}~\bibnamefont{Brofferio}},
  \bibinfo{author}{\bibfnamefont{O.}~\bibnamefont{Cremonesi}},
  \bibinfo{author}{\bibfnamefont{L.}~\bibnamefont{Gironi}},
  \bibinfo{author}{\bibfnamefont{M.}~\bibnamefont{Pavan}},
  \bibinfo{author}{\bibfnamefont{G.}~\bibnamefont{Pessina}},
  \bibinfo{author}{\bibfnamefont{S.}~\bibnamefont{Pirro}}, \bibnamefont{and}
  \bibinfo{author}{\bibfnamefont{E.}~\bibnamefont{Previtali}},
  \bibinfo{journal}{Astropart. Phys.} \textbf{\bibinfo{volume}{34}},
  \bibinfo{pages}{797 } (\bibinfo{year}{2011}).

\bibitem[{\citenamefont{Mikhailik et~al.}(2007)\citenamefont{Mikhailik, Henry,
  Kraus, and Solskii}}]{CMO_decay}
\bibinfo{author}{\bibfnamefont{V.~B.} \bibnamefont{Mikhailik}},
  \bibinfo{author}{\bibfnamefont{S.}~\bibnamefont{Henry}},
  \bibinfo{author}{\bibfnamefont{H.}~\bibnamefont{Kraus}}, \bibnamefont{and}
  \bibinfo{author}{\bibfnamefont{I.}~\bibnamefont{Solskii}},
  \bibinfo{journal}{Nucl. Instr. and Mech. A} \textbf{\bibinfo{volume}{583}},
  \bibinfo{pages}{350 } (\bibinfo{year}{2007}).

\bibitem[{\citenamefont{Mikhailik and Kraus}(2010)}]{Mikhailik_2010}
\bibinfo{author}{\bibfnamefont{V.~B.} \bibnamefont{Mikhailik}}
  \bibnamefont{and} \bibinfo{author}{\bibfnamefont{H.}~\bibnamefont{Kraus}},
  \bibinfo{journal}{Phys. Status Solidi B} \textbf{\bibinfo{volume}{247}},
  \bibinfo{pages}{1583} (\bibinfo{year}{2010}).

\bibitem[{\citenamefont{Annenkov et~al.}(2008)\citenamefont{Annenkov, Buzanov,
  Danevich, Georgadze, Kim, Kim, Kim, Kobychev, Kornoukhov, Korzhik
  et~al.}}]{Annenkov_CMO_property}
\bibinfo{author}{\bibfnamefont{A.~N.} \bibnamefont{Annenkov}},
  \bibinfo{author}{\bibfnamefont{O.~A.} \bibnamefont{Buzanov}},
  \bibinfo{author}{\bibfnamefont{F.~A.} \bibnamefont{Danevich}},
  \bibinfo{author}{\bibfnamefont{A.~S.} \bibnamefont{Georgadze}},
  \bibinfo{author}{\bibfnamefont{S.~K.} \bibnamefont{Kim}},
  \bibinfo{author}{\bibfnamefont{H.~J.} \bibnamefont{Kim}},
  \bibinfo{author}{\bibfnamefont{Y.~D.} \bibnamefont{Kim}},
  \bibinfo{author}{\bibfnamefont{V.~V.} \bibnamefont{Kobychev}},
  \bibinfo{author}{\bibfnamefont{V.~N.} \bibnamefont{Kornoukhov}},
  \bibinfo{author}{\bibfnamefont{M.}~\bibnamefont{Korzhik}},
  \bibnamefont{et~al.}, \bibinfo{journal}{Nucl. Instr. and Mech. A}
  \textbf{\bibinfo{volume}{584}}, \bibinfo{pages}{334 } (\bibinfo{year}{2008}).

\bibitem[{\citenamefont{Yuryev et~al.}(2011)\citenamefont{Yuryev, Jang, Kim,
  Lee, Lee, Lee, Yoon, and Kim}}]{yuryev_nim}
\bibinfo{author}{\bibfnamefont{Y.~N.} \bibnamefont{Yuryev}},
  \bibinfo{author}{\bibfnamefont{Y.~S.} \bibnamefont{Jang}},
  \bibinfo{author}{\bibfnamefont{S.~K.} \bibnamefont{Kim}},
  \bibinfo{author}{\bibfnamefont{K.~B.} \bibnamefont{Lee}},
  \bibinfo{author}{\bibfnamefont{M.~K.} \bibnamefont{Lee}},
  \bibinfo{author}{\bibfnamefont{S.~J.} \bibnamefont{Lee}},
  \bibinfo{author}{\bibfnamefont{W.~S.} \bibnamefont{Yoon}}, \bibnamefont{and}
  \bibinfo{author}{\bibfnamefont{Y.~H.} \bibnamefont{Kim}},
  \bibinfo{journal}{Nucl. Instr. and Mech. A} \textbf{\bibinfo{volume}{635}},
  \bibinfo{pages}{82 } (\bibinfo{year}{2011}).

\bibitem[{\citenamefont{Chernyak et~al.}(2012)\citenamefont{Chernyak, Danevich,
  Giuliani, Olivieri, Tenconi, and Tretyak}}]{chernyak_random_coin_dbd}
\bibinfo{author}{\bibfnamefont{D.~M.} \bibnamefont{Chernyak}},
  \bibinfo{author}{\bibfnamefont{F.~A.} \bibnamefont{Danevich}},
  \bibinfo{author}{\bibfnamefont{A.}~\bibnamefont{Giuliani}},
  \bibinfo{author}{\bibfnamefont{E.}~\bibnamefont{Olivieri}},
  \bibinfo{author}{\bibfnamefont{M.}~\bibnamefont{Tenconi}}, \bibnamefont{and}
  \bibinfo{author}{\bibfnamefont{V.~I.} \bibnamefont{Tretyak}},
  \bibinfo{journal}{Eur. Phys. J. C} \textbf{\bibinfo{volume}{72}}
  (\bibinfo{year}{2012}).

\bibitem[{\citenamefont{Artusa et~al.}(2014)\citenamefont{Artusa, Avignone~III,
  Azzolini, Balata, Banks, Bari, Beeman, Bellini, Bersani, Biassoni
  et~al.}}]{artusa2014exploring}
\bibinfo{author}{\bibfnamefont{D.}~\bibnamefont{Artusa}},
  \bibinfo{author}{\bibfnamefont{F.}~\bibnamefont{Avignone~III}},
  \bibinfo{author}{\bibfnamefont{O.}~\bibnamefont{Azzolini}},
  \bibinfo{author}{\bibfnamefont{M.}~\bibnamefont{Balata}},
  \bibinfo{author}{\bibfnamefont{T.}~\bibnamefont{Banks}},
  \bibinfo{author}{\bibfnamefont{G.}~\bibnamefont{Bari}},
  \bibinfo{author}{\bibfnamefont{J.}~\bibnamefont{Beeman}},
  \bibinfo{author}{\bibfnamefont{F.}~\bibnamefont{Bellini}},
  \bibinfo{author}{\bibfnamefont{A.}~\bibnamefont{Bersani}},
  \bibinfo{author}{\bibfnamefont{M.}~\bibnamefont{Biassoni}},
  \bibnamefont{et~al.}, \bibinfo{journal}{arXiv preprint arXiv:1404.4469}
  (\bibinfo{year}{2014}).

\bibitem[{\citenamefont{Chernyak et~al.}(2014)\citenamefont{Chernyak, Danevich,
  Giuliani, Mancuso, Nones, Olivieri, Tenconi, and
  Tretyak}}]{chernyak_rej_random_coin}
\bibinfo{author}{\bibfnamefont{D.~M.} \bibnamefont{Chernyak}},
  \bibinfo{author}{\bibfnamefont{F.~A.} \bibnamefont{Danevich}},
  \bibinfo{author}{\bibfnamefont{A.}~\bibnamefont{Giuliani}},
  \bibinfo{author}{\bibfnamefont{M.}~\bibnamefont{Mancuso}},
  \bibinfo{author}{\bibfnamefont{C.}~\bibnamefont{Nones}},
  \bibinfo{author}{\bibfnamefont{E.}~\bibnamefont{Olivieri}},
  \bibinfo{author}{\bibfnamefont{M.}~\bibnamefont{Tenconi}}, \bibnamefont{and}
  \bibinfo{author}{\bibfnamefont{V.~I.} \bibnamefont{Tretyak}},
  \bibinfo{journal}{Eur. Phys. J. C} \textbf{\bibinfo{volume}{74}},
  \bibinfo{eid}{2913} (\bibinfo{year}{2014}).

\bibitem[{\citenamefont{Lee et~al.}(2015)\citenamefont{Lee, So, Kang, Kim, Kim,
  Lee, Lee, Yoon, and Kim}}]{hjlee2015}
\bibinfo{author}{\bibfnamefont{H.~J.} \bibnamefont{Lee}},
  \bibinfo{author}{\bibfnamefont{J.~H.} \bibnamefont{So}},
  \bibinfo{author}{\bibfnamefont{C.~S.} \bibnamefont{Kang}},
  \bibinfo{author}{\bibfnamefont{G.~B.} \bibnamefont{Kim}},
  \bibinfo{author}{\bibfnamefont{S.~R.} \bibnamefont{Kim}},
  \bibinfo{author}{\bibfnamefont{J.~H.} \bibnamefont{Lee}},
  \bibinfo{author}{\bibfnamefont{M.~K.} \bibnamefont{Lee}},
  \bibinfo{author}{\bibfnamefont{W.~S.} \bibnamefont{Yoon}}, \bibnamefont{and}
  \bibinfo{author}{\bibfnamefont{Y.~H.} \bibnamefont{Kim}},
  \bibinfo{journal}{Nucl. Instr. and Meth. A} \textbf{\bibinfo{volume}{784}},
  \bibinfo{pages}{508} (\bibinfo{year}{2015}).

\end{thebibliography}

\end{document}